%% file: Libra.tex
\documentclass[final,12pt]{elsarticle}

\usepackage[utf8]{inputenc}
\usepackage[T1]{fontenc}

\usepackage{amssymb,amsmath}

\usepackage{mmacells}

\mmaDefineMathReplacement[≤]{<=}{\leq}
\mmaDefineMathReplacement[≥]{>=}{\geq}
\mmaDefineMathReplacement[≠]{!=}{\neq}
\mmaDefineMathReplacement[→]{->}{\to}[2]
\mmaDefineMathReplacement[⧴]{:>}{:\hspace{-.2em}\to}[2]
\mmaDefineMathReplacement{∉}{\notin}
\mmaDefineMathReplacement{∞}{\infty}
\mmaDefineMathReplacement{ϵ}{\mmaUnd{\epsilon}}
\mmaDefineMathReplacement{`}{\textasciigrave}

\mmaSet{
    leftmargin=.5em,
    morefv={gobble=4},
    morelst={alsoletter={`}},
    morecellgraphics={yoffset=1ex,xoffset=0}
}
\newcounter{bla}

\usepackage{xspace}
\newcommand{\inline}[2]{\mmaInlineCell[moredefined={History,ds1}]{Input}{History[ds1]}}

\newcommand{\Mathematica}{\textit{Mathematica}\xspace}

\newcommand{\Libra}{\texttt{Libra}\xspace}

\newcommand{\e}{\ensuremath{\epsilon}}
\newcommand{\bs}{\boldsymbol}
\newcommand{\Pexp}{\operatorname{Pexp}}

\newcommand{\sU}{\mathcal{U}}

\newcommand{\sV}{\mathcal{V}^\intercal}

\DeclareMathOperator{\im}{Im}
\DeclareMathOperator{\coim}{coIm}
\DeclareMathOperator{\coker}{coker}

\journal{Computer Physics Communications}

\begin{document}

\begin{frontmatter}



\title{\Libra: a package for transformation of differential 
systems for multiloop integrals.}


\author{Roman N. Lee\corref{author}}

\cortext[author] {\textit{E-mail address:} r.n.lee@inp.nsk.su}
\address{Budker Institute of Nuclear Physics,\\630090, Novosibirsk, Russia}
\begin{abstract}
We present a new package for \Mathematica system, called \Libra. Its purpose is to provide convenient tools for the transformation of the first-order differential systems $\partial_i \bs j = M_i \bs j$ for one or several variables. In particular, \Libra is designed for the reduction to $\e$-form  of the differential systems  which appear in multiloop calculations. The package also contains some tools for the construction of general solution: both via perturbative expansion of path-ordered exponent and via generalized power series expansion near regular singular points. \Libra also has tools to determine the minimal list of coefficients in the asymptotics of the original master integrals, sufficient for fixing the boundary conditions.
\end{abstract}


\end{frontmatter}



{\bf PROGRAM SUMMARY}

\begin{small}
\noindent
{\em Program title:} \texttt{Libra}  \\
{\em CPC Library link to program files:} (to be added by Technical Editor) \\
{\em Developer's respository link:} (if available) \\
{\em Code Ocean capsule:} (to be added by Technical Editor)\\
{\em Licensing provisions(please choose one):} GPLv3\\
{\em Programming language:} Wolfram Mathematica\\
{\em Nature of problem:}\\
Transformation of the first-order linear differential systems with respect to one or several variables: reduction to fuchsian form, to $\epsilon$-form, formal solution in terms of the Goncharov's polylogarithms and in terms of generalized power series.\\
{\em Solution method:}\\
Algorithms described in Refs. \cite{Lee2014,Lee2017c}, and \cite[Section E.8]{Blondel:2018mad}.

\end{small}

\section{Introduction}
\label{sec:intro}
\input{intro}

\section{Reduction to $\e$-form}
\input{reduction}

\section{\Libra package}\label{sec:libra}

\subsection{Package installation}
The \Libra package can be retrieved from its web site \href{https://rnlee.bitbucket.io/Libra/}{rnlee.bitbucket.io/Libra/} or directly from the bitbucket repository  \href{https://bitbucket.org/rnlee/libra/}{bitbucket.org/rnlee/libra/}. The installation amounts to unpacking the archive into any desired directory and to creating a shortcut file \texttt{init.m} in \textit{Mathematica} \texttt{\$UserBaseDirectory}. To automatize the creation of the shortcut, the \textit{Mathematica} script \texttt{makeShortcut.m} should be run. More detailed instructions can be found in the \texttt{INSTALL} text file.

After the installation, the package is loaded with the command \lstset{basicstyle=\small\sffamily}
\begin{mmaCell}[moredefined=Libra]{Input}
    <\!\!<Libra\textasciigrave
\end{mmaCell}
\vspace{-1cm}
\begin{center}
    \includegraphics[width=1\linewidth]{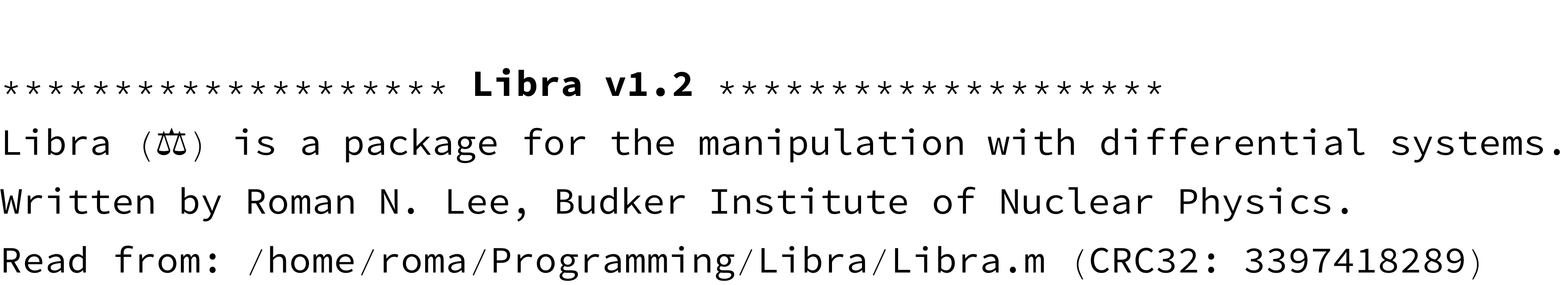}
\end{center}
Let us now present two examples of \Libra's usage.


\input{example1}
%
\input{example2}

\input{conclusion}

\bibliographystyle{elsarticle-num}
\bibliography{Libra}







\end{document}

%% file: intro.tex

Modern multiloop calculations essentially rely on the differential equations method. Within this approach, the integration-by-part (IBP) reduction is used to reduce all requred integrals to a finite set of master integrals and to construct the first-order differential systems for the latter. The possibility to find analytic solutions of these systems essentially relies on their reduction to some kind of canonical form. This may include, for example, the elimination of spurious singularities, reduction to local/global Fuchsian form, variable change and switching to/from one higher-order differential equation. Given high complexity of the differential systems which emerge in multiloop calculations, one can not avoid using computers for the reduction.

The present paper introduces a new package for \Mathematica system, called \Libra. Its purpose is to provide convenient tools for the transformation of the first-order differential systems.  The package also contains some tools for the construction of general solution: both via perturbative expansion of path-ordered exponent and via generalized power series expansion near regular singular points. \Libra also can help the user to determine the minimal list of coefficients in the asymptotics of the original master integrals near a singular point. Calculating the coefficients from this list is sufficient for fixing the boundary conditions.

One of the  most important tasks that \Libra can be relied on is the reducing of the differential equations to $\e$-form \cite{Henn2013,Lee2014}. In the next Section we review the corresponding reduction algorithm as presented in Refs. \cite{Lee2014,Lee2017c,Blondel:2018mad}. Note that this algorithm has been implemented in two publicly available codes, \texttt{epsilon}, Ref. \cite{Prausa2017a}, and \texttt{Fuchsia}, Ref. \cite{Gituliar2017}. However, we believe that \Libra will still be appreciated by the community due to its rich functionality and also high computational power. In this context, we note that \Libra has been already used in several bleeding-edge multiloop computations, see, e.g., Refs. \cite{Lee:2019zop,Grozin:2020jvt}. We describe the \Libra package in Section \ref{sec:libra}, where one can find two examples of \Libra usage.

The last, but not the least, although inspired by the multiloop applications, \Libra package is quite universal and can be used in other research fields, which require manipulations with the first-order differential systems of the form $\frac{\partial }{\partial x_i}\bs j = M_i(\bs x) \bs j$. In particular, these systems appear in algebraic geometry when considering Gauss-Manin connection.

%% file: reduction.tex
In the present section we shortly review the algorithm presented in Refs. 
\cite{Lee2014,Lee2017c,Blondel:2018mad}. We skip the description of 
IBP reduction method, Refs. \cite{Tkachov1981,ChetTka1981}, which is essential to obtain the 
differential equations for master integrals, \cite{Kotikov1991b,Remiddi1997}. 
We start from the differential system of the form 
\begin{equation}\label{eq:DEj}
    \frac{\partial}{\partial x_i}   \bs{j} =M_i\left(\bs x,\e\right)\bs j\qquad 
    (i=1,\ldots, N) . 
\end{equation}
Here $\bs j = (j_1,\ldots,j_n)^\intercal$ is a column of unknown functions, 
$\bs x =(x_1,\ldots x_N)$ are the variables, and $\e$ is a parameter. For typical multiloop calculation setup, these are the ``Laporta'' master integrals, the kinematic  invariants, and the dimensional regularization parameter, respectively. The matrices $M_i$ in the right-hand side rationally depend on $\bs x$ and $\e$. The observation made in Ref. \cite{Henn2013} is that by a proper choice of new functions $\bs J$ the differential system can often be cast in 
the form where the dependence on $\e$ is factorized in the right-hand side,\footnote{Note that the   observation of Ref. \cite{Henn2013} comes as surprise since the systems, reducible to $\e$-form, constitute a ``zero measure'' set among all systems with rational coefficients depending on parameter $\e$.}
\begin{equation}\label{eq:DEJ}
    \frac{\partial}{\partial x_i}   \bs{J} =\e S_i\left(\bs x\right)\bs J\qquad 
    (i=1,\ldots, N).
\end{equation}
In what follows we will refer to this form as \textit{$\e$-form}. Note that the integrability conditions applied to the system \eqref{eq:DEJ} imply that $M=\sum_i dx_i S_i(\bs x)$ is an exact 1-form (total differential) \cite{Henn2013}. 

The advantage of the differential system in $\e$-form is that its general solution
\begin{equation}
    U=\Pexp\left[\e \int d\bs x \cdot \bs S \right]
\end{equation}
can be readily expanded in $\e$-series in terms of iterated integrals. In particular, for rational matrices $S_i$ these integrals are nothing but the multiple polylogarithms \cite{Goncharov1998}.

Therefore, a natural question arises: given the differential system 
\eqref{eq:DEj} is it possible and how to find the transformation of functions 
$\bs j=T\bs J$, such that the new functions $\bs J$ satisfy the differential 
system \eqref{eq:DEJ}? Let us remark that, when answering to this question, it is important to restrict the class of transformations we want to consider. Otherwise, we can
always pass to trivial system $d\bs J =0$ by means of the formal transformation 
$T=\Pexp[\int d\bs x \cdot \bs M]=\Pexp[\int \sum_idx_i M_i]$. Our base case 
will be the class of transformations rational in both $\bs x$ and $\e$. At some point we will also consider the extension of this class to the transformations rational in some notations $\bs y$ algebraically related to the original variables via $x_i=f_i(\bs y)$ with $f_i$ being the rational 
functions\footnote{Since $f_i$ here are rational functions, the transformations 
which are rational in $\bs x$, are necessarily rational in $\bs y$. The 
inverse is, in general, not true, so this is indeed an extension of the class of transformations.}.

An algorithm of the reduction of a univariate differential system to $\e$-form has been 
suggested in Ref. \cite{Lee2014}. Later on, a strict criterion of irreducibility has been obtained in Ref. \cite{Lee2017c}. Also, in the same paper it was explained how to use the algorithm for multivariate setup. We will present now a basis-independent variant of this algorithm, partly described in Ref. \cite[Section E.8]{Blondel:2018mad}.

\subsection{Conventions and notations}
Note that the transformation of functions $\bs j=T\tilde{\bs j}$ is understood below as the transformation of matrices in the right-hand side of the differential system \eqref{eq:DEj}:
\begin{equation}
    M_i\to\widetilde{M}_i=T ^{-1}\left[M_i\,T-
    \frac{\partial}{\partial x_i} T\right].
\end{equation}

As explained in Ref. \cite{Lee2017c}, the problem of reduction to $\e$-form of the systems in several variables is effectively reduced to that of the system in one variable. One should simply proceed on one-by-one basis, reducing first the system in the first variable, then in the second variable, etc. The only restrictions are that the transformations considered for each successive variable should not depend on the previous variables. These restrictions seem to be easily fulfilled in every specific example (see, in particular, example 5 in \texttt{Tutorial1.nb} notebook attached to the distribution). 
Therefore, from now on we consider the differential equation in one variable.
\begin{equation}\label{eq:DE}
    \partial_x\bs j = M\bs j.
\end{equation}

Let us remind some facts from linear algebra and give a few useful definitions.  First, since we are aimed at a basis-independent treatment, we find it convenient to use a mixed terminology, coming partly from the operator language and partly from the matrix language. In particular, we have the dictionary presented in Table \ref{tab:dict}.

\begin{table}
    \centering
    \begin{tabular}{|rcl|}
        \hline
        Operator term&\vline&Matrix term\\
        \hline
        linear operator $L$ &$\leftrightarrow$& $n\times n$ matrix $L$\\\hline
        vector $\bs u$ &$\leftrightarrow$& column $\bs u =(u_1,\ldots,u_n)^\intercal$\\\hline
        covector $\bs v^\intercal$ &$\leftrightarrow$& row $\bs v^\intercal=(v_1,\ldots,v_n)$\\\hline
        $\ker L$ &$\leftrightarrow$& $\{\bs u ,\, L\bs u=0\}$\\\hline
        $\coker L$ &$\leftrightarrow$& $\{\bs v^\intercal ,\,\bs v^\intercal L=0\}$\\\hline
        $\im L$ &$\leftrightarrow$& $\{\bs u ,\, \exists \tilde{\bs u}: L\tilde{\bs u} = \bs u\}$\\\hline
        $\coim L$ &$\leftrightarrow$& $\{\bs v^\intercal ,\,\exists \tilde{\bs v}^\intercal :\tilde{\bs v}^\intercal L=\bs v^\intercal\}$\\\hline
    \end{tabular}
    \caption{Translation dictionary between the operator and matrix languages.}\label{tab:dict}
\end{table}

Moreover, we find it convenient to identify the $k$-dimensional linear subspace (or simply $k$-subspace) $\mathcal U$ with the $n\times k$ matrix whose columns form some basis of this subspace. Although this matrix is not uniquely defined, nevertheless, every expression below that contains this matrix will be independent on the specific choice of the basis, so we denote this matrix also as  $\mathcal U$. For example, the statement $\mathcal{U}\subset \ker L$ is equivalent to $L\mathcal{U}=0$. In what follows, the invariant subspaces of operators will appear. Within our mixed language the wording ``$\mathcal U$ is invariant subspace of $L$'' is equivalent to ``$\exists$ matrix $C$ such that  $L \mathcal U =\mathcal U C$''.

Also, we will refer to the element $\bs v^\intercal$ and subspace $\mathcal{V}^\intercal$ in the dual space as \textit{right} vector and \textit{right} subspace, as opposed to the \textit{left} vector $\bs u$ and \textit{left} subspace $\mathcal{U}$ for the original space\footnote{I.e., left vectors can be mutiplied by a matrix from the left, and vice versa.}. Note that the matrix, corresponding to $k$-dimensional right subspace $\mathcal{V}^\intercal$ has dimension $k\times n$ (rather than $n\times k$) which should be easy to memorize thanks to $\bullet^\intercal$ notation.

Finally, let us fix our notations for describing the properties of the differential system \eqref{eq:DE} in the vicinity of some point $x_0$. Let $M(x)$ have the following series expansion at $x=x_0\neq \infty$:
\begin{equation}
    M(x)=\sum_{k=-r-1}^{\infty} M_k\cdot  (x-x_0)^k\,,
\end{equation}
where $M_{-r-1}\neq 0$.
Then we  say that $\max(r,0)$ is a \textit{Poincare rank at $x=x_0$} and $M_{-1}$ is a \textit{matrix residue at $x=x_0$}. If $r\geqslant 0$ ($r<0$), we say that $x_0$ is a singular point (regular point).
Similarly, for the expansion near $x=\infty$,
\begin{equation}
    M(x)=\sum_{k=-r+1}^{\infty} M_k\cdot  x^{-k}\,,
\end{equation}
we  say that $\max(r,0)$ is a \textit{Poincare rank at $x=\infty$} and $-M_{1}$ is a \textit{matrix residue at $x=\infty$ (note the minus sign)}. If $r\geqslant 0$ ($r<0$), we say that $\infty$ is a singular point (regular point).

\subsection{Projector and balance transformation}

As is well-known, the projector operator $P=P^2$ is totally defined by its image and kernel, or, equivalently, by its image and co-image, $\mathcal{V}^\intercal =\coim P$ and $\mathcal{U} =\im P$. Given the left subspace $\mathcal{V}^\intercal$ and the right subspace $\mathcal{U}$ of equal dimension $k$, one can construct a projector
\begin{equation}\label{eq:Pr}
    P=P(\sU,\sV)=\sU(\sV\sU)^{-1}\sV\,.
\end{equation}
This is true unless the $k\times k$ matrix $\sV\sU$ is not invertible.
Note that the matrix in the right-hand side of Eq. \eqref{eq:Pr} is defined uniquely despite the freedom of choice of $\sU$ and $\sV$. The properties $P^2=P,\ \im P=\sU,\ \coim P = \sV$ can be readily verified.

Given a projector $P$, we define \textit{$P$-balance transformation between two finite points $x_1$ and $x_2$} as 
\begin{equation}
    B(P,x_1,x_2|x)=\overline{P}+\frac{x-x_2}{x-x_1}P, 
\end{equation}
where $\overline{P}=1-P$. Similarly, we define
\begin{equation}
    B(P,x_1,\infty|x)=\overline{P}+\frac{1}{x-x_1}P,\quad 
    B(P,\infty,x_2|x)=\overline{P}+ (x-x_2)P.
\end{equation}
When $P=P(\sU,\sV)$ we also write $B(\sU,\sV|x_1,x_2|x) = B(P(\sU,\sV),x_1,x_2|x)$.

The balance transformation may change the Poincare rank of $M$ and the eigenvalues of the matrix residue in two points, $x=x_1$ and $x=x_2$, at most.

We will call the balance transformation $B(\sU,\sV|x_1,x_2|x)$ an \textit{$x_1$-adjusted} (\textit{$x_2$-adjusted}) if it does not increase the Poincare rank of the system at $x=x_1$ (at $x=x_1$). It is easy to prove that $B(\sU,\sV|x_1,x_2|x)$ is \textit{$x_1$-adjusted} iff $x_1$ is a singular point and $\sU$ is a left invariant subspace of the leading series expansion coefficient of $M(x)$ near $x=x_1$. Similarly, $B(\sU,\sV|x_1,x_2|x)$ is \textit{$x_2$-adjusted} iff $x_2$ is a singular point and $\sV$ is a right invariant subspace of the leading series expansion coefficient near $x=x_2$.

In what follows we will always use the balances that are adjusted in both points, unless otherwise stated.

\subsection{Reducing Poincare rank and shifting eigenvalues of matrix residues}
Let us consider now the point of positive Poincare rank, e.g., let it be $x=0$. So, we have ($r>0$)
\begin{equation}
    M(x)=\sum_{k=-r-1}^{\infty} M_k\cdot  x^k = \frac{A_0}{x^{r+1}}+\frac{A_1}{x^{r}}+\ldots\,.
\end{equation}
We want to find the adjusted balance transformation to strictly reduce the matrix rank of $A_0=M_{-r-1}$. Let us search for the balance having zero as the first singular point,  $B(\sU,\sV|0,x_2|x)$. As it is described in Ref. \cite[Section E.8]{Blondel:2018mad}, it suffices to find $\sU$ with the following properties:
\begin{equation}
    \mathrm{I.}\quad  A_{0}\sU=0,\qquad
    \mathrm{II.}\quad  A_{1}\sU\subseteq \im A_{0}+\sU,\qquad
    \mathrm{III.}\quad  \sU\cap\im A_{0}>\{0\}.\label{eq:conds}
\end{equation}
A necessary and sufficient criterion of the existence of such a $\sU$ is \cite[Section E.8]{Blondel:2018mad}
\begin{gather}\label{eq:crit2}
    \dim \ker \mathcal{A} > \dim \ker A_0\,,\nonumber
\intertext{where}
    \mathcal{A}=\begin{pmatrix}A_{0} & A_{1}-\lambda\\ 0 & A_{0}\end{pmatrix}\,.
\end{gather}
This condition simply states that the number of null eigenvectors of the matrix $\mathcal{A}$ should be greater than that of $A_0$. Let us note that if $\bs u$ is a null eigenvector of $A_0$ then $\bs u \choose \bs 0 $ is that of $\mathcal{A}$, and \emph{vice versa}. Therefore, the criterion \eqref{eq:crit2} states that there must be at least one null eigenvector of $\mathcal{A}$ of the form $\bs w(\lambda)\choose \bs u(\lambda)$ with $\bs u(\lambda)\neq0$. The null eigenvectors can be found routinely and their components are, in general, the rational functions of $\lambda$. As one can always get rid of common denominator, we can assume that these components are polynomials in $\lambda$ and
\begin{equation}\label{eq:ulambda}
    \bs u(\lambda)=\bs u_0+\bs u_1\lambda +\ldots \bs u_k \lambda^k
\end{equation}
Now, we can easily check that $\sU=\mathrm{span}(\bs u_0,\ldots ,\bs u_k)$ satisfies conditions \eqref{eq:conds}.

If we are looking for the balance  $B(\sU,\sV|x_1,0|x)$, having zero as the second singular point, we have some requirements for $\sV$. To construct $\sV$ with the required properties, we find the \textit{right} null eigenvector $(\bs v^\intercal(\lambda),\bs w^\intercal(\lambda))$ of $\mathcal{A}$, such that 
\begin{equation}\label{eq:vlambda}
    v^\intercal(\lambda)=\bs v^\intercal_0+\bs v^\intercal_1\lambda +\ldots \bs v^\intercal_k \lambda^k \neq \bs 0\,.
\end{equation} Then we put $\sV=\mathrm{span}(\bs v^\intercal_0,\ldots ,\bs v^\intercal_k)$. 

Now, let the system have zero Poincare rank (Fuchsian singularity) at $x=0$:
\begin{equation}
    M(x)=\sum_{k=-1}^{\infty} M_k\cdot  x^k = \frac{A_0}{x}+A_1+\ldots\,.
\end{equation}
Let us consider the balance $B(\sU,\sV|0,x_2|x)$, where $\sU$ is the left invariant subspace of $A_0$. In a suitable basis the matrices $P=P(\sU,\sV)$, $A_0$, and $A_1$ have the following block forms
\begin{equation}
    P=\begin{pmatrix}1&0\\0&0\end{pmatrix}\,,\quad A_0=\begin{pmatrix}B_1&B_2\\0&B_3\end{pmatrix}
    \,,\quad A_1=\begin{pmatrix}B_4&B_5\\B_6&B_7\end{pmatrix}
\end{equation}
Then the expansion of the transformed matrix starts from $\frac{\tilde{A}_0}{x}$ where
\begin{equation}
    \tilde{A}_0=\overline{P}A_0+(A_0+1)P+\overline{P}A_1P=\begin{pmatrix}B_1+1&0\\B_6&B_3\end{pmatrix}
\end{equation}
Thus, the eigenvalues, corresponding to $\sU$ are shifted by $+1$.

If we instead apply the balance $B(\sU,\sV|x_1,0|x)$ with $\sV$ being the right-invariant subspace of $A_0$, the corresponding eigenvalues are shifted by $-1$.

Let us summarize the content of this subsection. We have shown that the balance transformation  $B(\sU,\sV|x_1,x_2|x)$ can be used to reduce the Poincare rank and to shift the eigenvalues at Fuchsian singular points. Moreover, the discussed properties of the transformation at $x=x_1$ depend only on $\sU$, while those  at $x=x_2$ depend on $\sV$, so these two subspaces can be constructed almost independently. The only requirement that simultaneously involves $\sU$ and $\sV$ is that $\sV\sU$ should be a square invertible matrix.

\subsection{Factoring out $\epsilon$}

Suppose now that we have been able to find the global Fuchsian form with all eigenvalues of matrix residues proportional to $\e$. It means that the matrix in the right-hand side now has the form
\begin{equation}
    M(x,\e)=\sum_k \frac{M_k(\e)}{x-x_k}
\end{equation}
and all eigenvalues of all $M_k(\e)$ are proportional to $\e$. Then, thanks to Proposition 1  from Ref. \cite{Lee2017c}, if $\e$-form exists, it can be obtained by some transformation $T(\e)$ independent of $x$. In order to find this transformation, we solve the linear system of equations \cite{Lee2014}
\begin{equation}\label{eq:factor_out}
    \frac{M_k(\e)}{\e}T(\e,\mu)=T(\e,\mu)\frac{M_k(\mu)}{\mu}
\end{equation}
with respect to the matrix elements of $T(\e,\mu)$.
Note that from Eq. \eqref{eq:factor_out} it follows that 
\begin{equation}\label{eq:factor_out1}
    \frac{M(x,\e)}{\e}T(\e,\mu)=T(\e,\mu)\frac{M(x,\mu)}{\mu}
\end{equation}  
If this system has the invertible solution for some $\mu$, the transformation $T(\e)=T(\e,\mu)$ reduces our system to $\e$-form, which becomes obvious once we multiply \eqref{eq:factor_out1} by $\e\, T^{-1}$ from the left. We note here, that in real-life examples $x_k$ might be complicated algebraic numbers not even expressible in terms of radicals. Then, instead of solving the system \eqref{eq:factor_out}, we might use sufficient number of rational sampling points $x = a_1,\ldots a_m$ and solve the linear system
\begin{equation}\label{eq:factor_out2}
    \frac{M(a_k,\e)}{\e}T(\e,\mu)=T(\e,\mu)\frac{M(a_k,\mu)}{\mu},\quad (k=1,\ldots,m),
\end{equation}
which has the advantage of having rational coefficients. Besides, this method can be easily generalized to multivariate case.

\subsection{Using the block-triangular form}

The real-life examples of the differential systems which one might meet in the contemporary multiloop calculations may include several hundreds of equations. It is neither possible nor necessary to reduce such systems as a whole. Rather, one should use the block-triangular structure of the systems as already described in Ref. \cite{Lee2014}. Here we will only note that reducing the powers of irreducible denominators in the off-diagonal elements can be done without factorizing them into linear factors. Suppose, e.g., that we have in our system some denominator 
$$d(x)=d_0+d_1 x+\ldots +d_n x^n$$
being the irreducible polynomial of $n$-th degree. Using the same notations as in Eq. (7.1) of Ref. \cite{Lee2014} we have the differential systems 
\begin{align}\label{eq:offdiagonal}
   d(x)\partial_x \bs J_1 &= \e A(x) \bs J_1 + \frac{B(x,\e)}{d(x)^r}\bs J_2+\ldots,\nonumber\\
   d(x)\partial_x \bs J_2 &= \e C(x) \bs J_2 +\ldots,
\end{align}
Here $r$ is the Poincare rank  of the system at any zero of $d(x)$ and the dots denote the terms which are either less singular at zeros of $d(x)$ or correspond to the contributions of lower sectors. By assumption $r>0$. Note that we may restrict ourselves to the case when the matrices $A(x)$, $B(x,\e)$ and $C(x)$ have entries which are polynomials of, at most, $(n-1)$-th degree. Indeed, suppose that some entry of $A$, $B$, or $C$ has the form $\frac{p(x)}{q(x)}$, where $p(x)$ and $q(x)$ are coprime polynomials and $q(x)$ is coprime with $d(x)$. Then we use a well-known technique to reduce the rational function $\frac{p(x)}{q(x)}$ with respect to $d(x)$. Thus, we have 
\begin{equation}
    \frac{p(x)}{q(x)}=r(x)+d(x)\frac{s(x)}{q(x)}\,,
\end{equation}
where $r(x)$ and $s(x)$ are some polynomials, and $r(x)$ has degree $n-1$ at most. Note that we basically have to treat $x$ as algebraic extension of $\mathbb{Q}$, defined by the equation $d(x)=0$. In particular, we use the extended Euclidean algorithm to invert $q(x)$. The corresponding procedures are implemented in many computer algebra systems, e.g., in \texttt{Fermat} \cite{lewis2008computer}, but, unfortunately, not in \texttt{Mathematica}. We have implemented the required algorithms in \Libra, so that $r(x)$ can be obtained by the command \mmaInlineCell[moredefined=QuolyMod]{Input}{QuolyMod[\(p(x)/q(x)\),\(x\to d(x)\)]}.
In what follows we will adopt the notation
\begin{equation}
    \frac{p(x)}{q(x)}=r(x)\quad (\mathrm{mod}\ d).
\end{equation}

Then we can replace $\frac{p(x)}{q(x)}$ with $r(x)$ and move the term $d(x)\frac{s(x)}{q(x)}$ to the part hidden with dots in Eq. \eqref{eq:offdiagonal}.
So, we assume that 
\begin{equation*}
    A(x)=\sum_{k=0}^{n-1} A_k x^k,\quad
    B(x,\e)=\sum_{k=0}^{n-1} B_k(\e) x^k,\quad
    C(x)=\sum_{k=0}^{n-1} C_k x^k.
\end{equation*}
Then we make the substitution 
\begin{equation}
    \bs J_1 =\tilde{\bs J}_1+d(x)^{-r} D \bs J_2,
\end{equation}
where $D=D(x,\e)=\sum_{k=0}^{n-1} D_k(\e) x^k$ is some matrix. Collecting the most singular terms, containing $d^{-r}$, we obtain
\begin{equation}
    d(x)\partial_x \tilde{\bs J}_1 = \e A \tilde{\bs J}_1 + \left[rd^\prime D+\e (AD-DC)+ B\right]d(x)^{-r}J_2+\ldots
\end{equation}
Then we have the equation 
\begin{equation}
    d^\prime D+\frac{\e}{r} (AD-DC)=-\frac{B}{r} \quad (\mathrm{mod}\ d)
\end{equation}
Since $d$ is irreducible, $d^\prime$ is coprime with $d$, so we can divide the equation by $d^\prime$,
\begin{equation}
    D+\frac{\e}{r} \frac{AD-DC}{d^\prime }=-\frac{B}{rd^\prime} \quad (\mathrm{mod}\ d),
\end{equation}
and think of $\frac{AD-DC}{d^\prime}$ and $\frac{B}{rd^\prime}$ as some polynomial matrices of degree $n-1$. Then this equation necessarily has a solution as the linear operator acting on $D$ in the right-hand side is close to unity for sufficiently small $\e$ and, thus, is invertible for generic $\e$.

%% file: example1.tex

\subsection{Example 1: Reducing system of 5 equations}
Let us consider the differential system of 5 equations with the following matrix standing in the right-hand side:
\lstset{basicstyle={\tiny\sffamily}}
{\begin{mmaCell}{Input}
    m =
\end{mmaCell}

\vspace{-0.993cm}

{\tiny
\begin{verbatim}
       {{(1-2*x+x^2-5*e+10*x*e-7*x^2*e+6*e^2-8*x*e^2+6*x^2*e^2)/((-1+x)*x*(1+x)*e),(-2*(-1+x))/(x^2*(1+x)),
    -2*(1-x)*(1-3*e)/(x^2*(1+x)*e),2*(1-x)/(x^2*(1+x)),2*(1-x)*(1+2*e)/(x^2*(1+x)*e)},{(1-3*e)*(1-4*e)*x/(1-x^2),
    (1+x^2-2*x*e)/((1-x^2)*x),(2*(1-5*e))/(1-x^2),-6*e/(1-x^2),0},{((1-3*e)*(1-4*e)*((1-x)^2-2*e-2*x^2*e))/(2*(1-x^2)*e),
    (-1+2*x-x^2+4*e-4*x*e+4*x^2*e)/((1-x^2)*x),(1-2*x+x^2-6*e+10*x*e-6*x^2*e+12*e^2-8*x*e^2+12*x^2*e^2)/((1-x^2)*x*e),
    (-(1-x)^2+4*e*(1-x+x^2))/((1-x^2)*x),-(1-x)*(1+2*e)*(1-3*e)/(x*(1+x)*e)},{(x*(1-3*e)*(1-4*e))/(1-x^2),-6*e/(1-x^2),
    (2*(1-5*e))/(1-x^2),(1+x^2-2*x*e)/((1-x^2)*x),0},{(-1+4*e)*(1+x+x^2-e*(5-x+5*x^2)+2e^2*(3-x+3*x^2))/((1-x^2)*(1+2*e)),
    2*e*(1+x+x^2-4*e+6*x*e-4*x^2*e)/((1-x^2)*x*(1+2*e)),-2*(1+x+x^2-e*(7+x+7*x^2)+2*e^2*(6-5*x+6*x^2))/((1-x^2)*x*(1+2*e)),
    2*e*(1+x+x^2-4*e+6*x*e-4*x^2*e)/((1-x^2)*x*(1+2*e)),2*(1+x^2-2*e+6*x*e-2*x^2*e)/((1-x^2)*x)}}/.e->\[Epsilon];
\end{verbatim}
}}
\lstset{basicstyle=\small\sffamily}
We initialize the new differential system with the command
\begin{mmaCell}[moredefined={NewDSystem, m}]{Input}
    NewDSystem[ds1,x\(\pmb{\to}\)m];
\end{mmaCell}
The effect of this command is that all following subsequent transformations will be attached to \mmaInlineCell{Code}{ds1} symbol. In particular, we can use \linebreak \mmaInlineCell[moredefined=ds1]{Code}{DumpSave["ds1.m", ds1]} to save our work to a file  \texttt{ds1.m} to recover later with  \mmaInlineCell{Code}{Get["ds1.m"]}. The original matrix \mmaInlineCell[moredefined=m]{Input}{m} can be printed with \mmaInlineCell[moredefined={ds1}]{Input}{ds1[x]}.
Now we are going to run the visual tool in order to find the appropriate transformation
\begin{mmaCell}[moredefined={VisTransformation,ds1}]{Input}
    t=VisTransformation[ds1,x,ϵ];
\end{mmaCell}
After invoking this command the following window appears:
\begin{figure}[h]
    \includegraphics[width=1\linewidth]{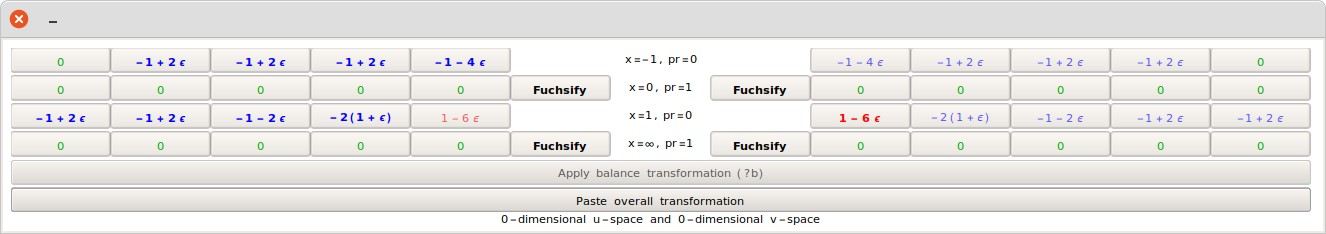}
    \caption{Visual tool for finding a (sequence of) balancing transformation(s).}
    \label{fig:VT1}
\end{figure}

This interface allows one to construct a balance transformation from the two subspaces, $\sU$ and $\sV$, which are defined using the left and the right halves of the window, respectively. When the two subspaces are suitable for constructing the projector, i.e., have equal nonzero dimension and the matrix $\sV\sU$ is invertible, the two lower buttons labeled by ``\mmaInlineCell{Print}{Apply balance transformation}'' (below referred to as ``\mmaInlineCell{Print}{Apply}'' button) and ``\mmaInlineCell{Print}{Paste overall transformation}'' (``\mmaInlineCell{Print}{Paste}'' button) turn green and enabled. When the matrix $\sV\sU$ is not invertible, those two buttons turn red. The effect of pressing the button ``\mmaInlineCell{Print}{Paste}'' is that the found transformation is returned (in our case, it is assigned to the variable \mmaInlineCell{Input}{t}). The button ``\mmaInlineCell{Print}{Apply}''  transforms the temporary matrix (which was first initialized by \mmaInlineCell[moredefined={VisTransformation,ds1}]{Input}{ds1[x]}) with constructed balance and recalculates the interface respectively. Later, when the button ``\mmaInlineCell{Print}{Paste}'' is pressed, the tool returns the product of all applied transformations.
Each row in this window, apart from the two lower rows occupied by wide buttons, corresponds to a singular point of the system, as indicated in the middle part of the interface. For example, ``\mmaInlineCell{Print}{x=-1, pr=0}'' corresponds to a singular point $x=-1$ with Poincare rank 0. Each button in the row, except those labeled ``\mmaInlineCell{Print}{Fuchsify}'', corresponds to the eigenvalue (which is shown on the button) of the leading series coefficient (for zero Poincare rank it is the matrix residue). When such a button is toggled down, the corresponding eigenvector is added to the basis of $\sU$ (left half) or $\sV$ (right half). The current dimensions of $\sU$ and $\sV$ are indicated in the status line. For the points with positive Poincare rank there is an additional button (in fact, there may be several buttons) ``\mmaInlineCell{Print}{Fuchsify}''. This button corresponds to the subspace spanned by vector $u(\lambda)$, Eq. \eqref{eq:ulambda}, or   $v^\intercal(\lambda)$, Eq. \eqref{eq:vlambda}, depending on whether the left or right half of the table is concerned. Pressing these buttons may increase the dimension of $\sU$ or $\sV$ by more than 1.

Let us now return to our example. 
\begin{enumerate}
\item We first want to reduce the Poincare at $x=0$ and at $x=\infty$.
We can do it in one step, by toggling the left ``\mmaInlineCell{Print}{Fuchsify}'' button in the second line and the right ``\mmaInlineCell{Print}{Fuchsify}'' button in the fourth line.
To avoid unnecessary repetitions of the window screenshots, let us  agree about the numbering of the buttons: in the left half of the table the buttons will be numbered in left-to-right top-to-bottom order, while in the right half they will be numbered in right-to-left top-to-bottom order. With this numbering we toggle button \#$11$ to the left and button \#$22$ to the right (see Fig. \ref{fig:VT1}). Then we press ``\mmaInlineCell{Print}{Apply}''. We will denote this sequence of actions as $\{11\}\longleftrightarrow\{22\}$. Then a new window appears:
\begin{center}
\includegraphics[width=0.9\linewidth]{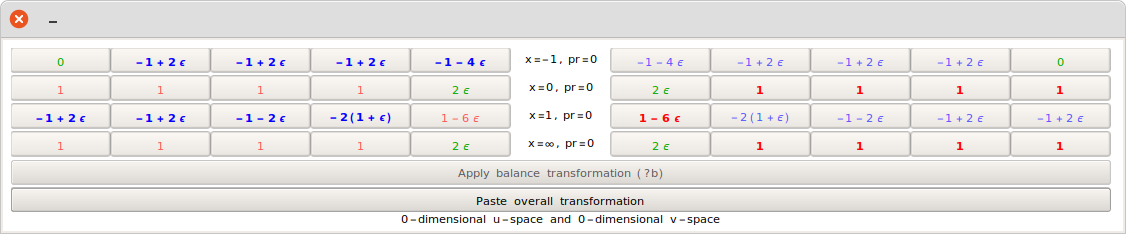}
\end{center}
\item We see that, indeed, the Poincare rank at $x=0,\,\infty$ has been reduced to zero, while the eigenvalues  at $x=\pm1$ remained intact. Now we can try to increase the four negative eigenvalues of the matrix residue at $x=-1$ and simultaneously increase the four positive ones at $x=0$. Thus, we toggle down buttons \#\#$2,3,4,5$ to the left, and buttons \#\#$6,7,8,9$ to the right. Then we press ``\mmaInlineCell{Print}{Apply}'' again.  In short notations, we apply
\begin{equation*}
    \{2,3,4,5\}\longleftrightarrow\{6,7,8,9\}\,
\end{equation*}
balance.
The result is
\begin{center}
    \includegraphics[width=0.9\linewidth]{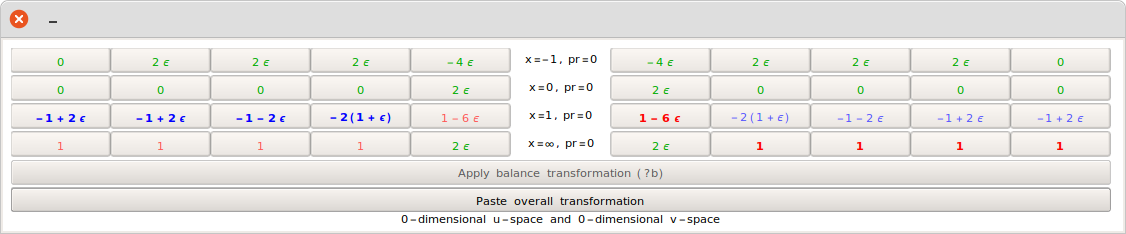}
\end{center}
We see that, indeed, we have managed to accomplish our goal: the eigenvalues of the matrix residues at $x=-1$ and $x=0$ are now all proportional to $\e$.
\item Similarly, we apply
\begin{equation*}
    \{11,12,13,14\}\longleftrightarrow\{16,17,18,19\}
\end{equation*}
and obtain
\begin{center}
    \includegraphics[width=0.9\linewidth]{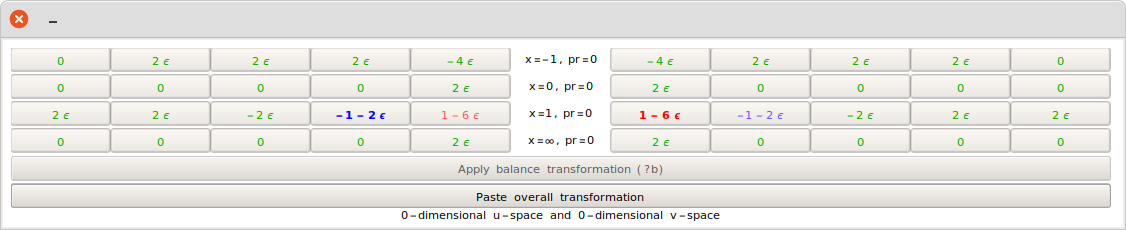}
\end{center}
\item At this stage, we have one negative and one positive eigenvalue at $x=1$. We can not balance them in one step. Therefore, we first ``move'' one of them to another point. E.g., we apply
\begin{equation*}
    \{14\}\longleftrightarrow \{20\}
\end{equation*}
and obtain
\begin{center}
\includegraphics[width=0.9\linewidth]{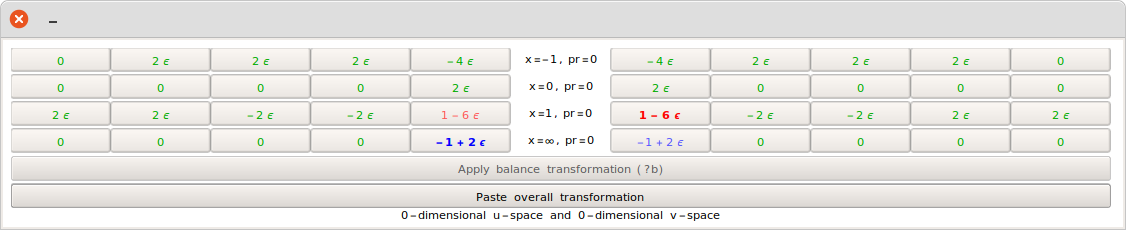}
\end{center}
\item Finally, applying
\begin{equation*}
    \{20\}\longleftrightarrow\{15\}
\end{equation*}
we have
\begin{center}
    \includegraphics[width=0.9\linewidth]{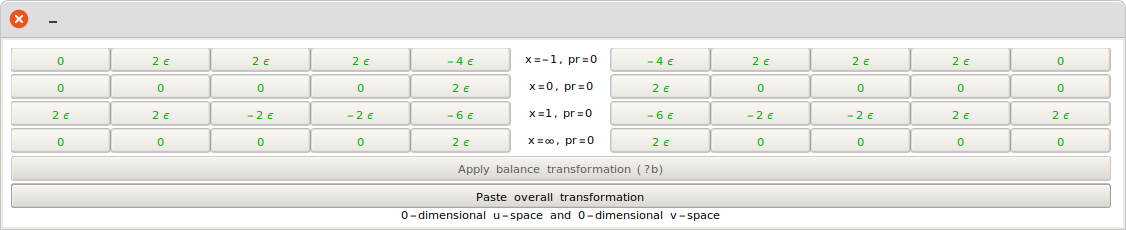}
\end{center}
\item 
At this stage we have reached global fuchsian form with all eigenvalues of all matrix residues proportional to $\e$. We press ``\mmaInlineCell{Print}{Paste}'' to assign the found transformation to the variable \mmaInlineCell{Input}{t}.
\end{enumerate}
Note that we did not change the differential system yet: 
\mmaInlineCell[moredefined={ds1, m}]{Input}{ds1[x]===m} will return \mmaInlineCell{Output}{True}.
In order to apply the transformation to  \mmaInlineCell[moredefined=ds1]{Input}{ds1} we execute
\begin{mmaCell}[moredefined={Transform,ds1,t}]{Input}
    Transform[ds1,t];
\end{mmaCell}
Now \mmaInlineCell[moredefined={m,ds1}]{Input}{ds1[x]===m} returns \mmaInlineCell{Output}{False}. Note that \mmaInlineCell[moredefined={Transform,ds1,t}]{Input}{Transform[ds1,t];} not only modifies the differential system, but it also ``registers'' the applied transformation in a special list \mmaInlineCell[moredefined={History,ds1}]{Input}{History[ds1]} associated with \mmaInlineCell[moredefined=ds1]{Input}{ds1}. This list spares the necessity to manually keep track of the applied transformations. Also, thanks to this list, we can easily undo one or several last transformations with \mmaInlineCell[moredefined={Undo,ds1}]{Input}{Undo[ds1]} or  \mmaInlineCell[moredefined={Undo,ds1,n}]{Input}{Undo[ds1,n]}.

Now we can find the constant transformation which factors out $\e$ with the command \mmaInlineCell[moredefined={t, FactorOut, ds1}]{Input}{FactorOut[ds1[x],x,\mmaUnd{\(\pmb{\epsilon}\)},\mmaUnd{\(\pmb{\mu}\)}]}. This command gives the most general matrix which satisfies linear system \eqref{eq:factor_out}. Apart from the parameter $\mu$, the output also depends on some unfixed constants of the form \texttt{C[}$k$\texttt{]}. Later on we have to put all those constants to some numbers generic enough so that $T$ remains invertible (assuming that it was invertible for unfixed constants). So, to save some space, we call the \mmaInlineCell[moredefined={t, FactorOut, ds1}]{Input}{FactorOut} function with $\mu=1$ and replace the remaining constants with some specific values, checking afterwards that the resulting matrix has non-zero determinant:
\begin{mmaCell}[moredefined={t, FactorOut, ds1,Factor,Det}]{Input}
    t=FactorOut[ds1[x],x,\mmaUnd{\(\pmb{\epsilon}\)},1]/.\(\pmb{\,}\)\{C[1]\(\pmb{\to}\)1,_C\(\pmb{\to}\)0\}; Factor[Det[t]]=!\!\! =0
\end{mmaCell}
\begin{mmaCell}{Output}
    True
\end{mmaCell}
In addition, putting $\mu$ to some number can accelerate the \mmaInlineCell[moredefined={t, FactorOut, ds1}]{Input}{FactorOut} procedure. Finally, we apply the found transformation
\begin{mmaCell}[moredefined={Transform,ds1,t}]{Input}
    Transform[ds1,t]
\end{mmaCell}
\lstset{basicstyle={\tiny\sffamily}}
\begin{mmaCell}{Output}
    \{\{-\mmaFrac{2 (-1-38 x+15 \mmaSup{x}{2}) \(\epsilon\)}{(x-1) x (1+x)},\mmaFrac{2 (1-7 x+3 \mmaSup{x}{2}) \(\epsilon\)}{3 (x-1) x (1+x)},\mmaFrac{4 (5-45 x+18 \mmaSup{x}{2}) \(\epsilon\)}{3 (x-1) x (1+x)},\mmaFrac{2 (1-7 x+3 \mmaSup{x}{2}) \(\epsilon\)}{3 (x-1) x (1+x)},-\mmaFrac{4 (-1-15 x+6 \mmaSup{x}{2}) \(\epsilon\)}{(x-1) x (1+x)}\},
    \{-\mmaFrac{6 (-8+5 x) \(\epsilon\)}{x (1+x)},\mmaFrac{2 (-2+x) \(\epsilon\)}{x (1+x)},\mmaFrac{8 (-5+3 x) \(\epsilon\)}{x (1+x)},\mmaFrac{2 (2 \(\epsilon\)-x \(\epsilon\)+\mmaSup{x}{2} \(\epsilon\))}{(x-1) x (1+x)},-\mmaFrac{12 (-3+2 x) \(\epsilon\)}{x (1+x)}\},\{\mmaFrac{9 \(\epsilon\)}{x},-\mmaFrac{\(\epsilon\)}{x},-\mmaFrac{2 (-5+3 \mmaSup{x}{2}) \(\epsilon\)}{(x-1) x (1+x)},-\mmaFrac{\(\epsilon\)}{x},\mmaFrac{6 \(\epsilon\)}{x}\},
    \{-\mmaFrac{6 (-8+5 x) \(\epsilon\)}{x (1+x)},\mmaFrac{2 (2 \(\epsilon\)-x \(\epsilon\)+\mmaSup{x}{2} \(\epsilon\))}{(x-1) x (1+x)},\mmaFrac{8 (-5+3 x) \(\epsilon\)}{x (1+x)},\mmaFrac{2 (-2+x) \(\epsilon\)}{x (1+x)},-\mmaFrac{12 (-3+2 x) \(\epsilon\)}{x (1+x)}\},
    \{\mmaFrac{13 (-2-7 x+3 \mmaSup{x}{2}) \(\epsilon\)}{(x-1) x (1+x)},-\mmaFrac{(-4-18 x+9 \mmaSup{x}{2}) \(\epsilon\)}{3 (x-1) x (1+x)},-\mmaFrac{2 (-20-107 x+45 \mmaSup{x}{2}) \(\epsilon\)}{3 (x-1) x (1+x)},-\mmaFrac{(-4-18 x+9 \mmaSup{x}{2}) \(\epsilon\)}{3 (x-1) x (1+x)},\mmaFrac{2 (-11-36 x+15 \mmaSup{x}{2}) \(\epsilon\)}{(x-1) x (1+x)}\}\}
\end{mmaCell}
Note that the differential system now is in $\e$-form. Let us try find a constant transformation (independent of $x$ and $\e$) which somewhat simplifies the numerical coefficients of the system. We might try to transform one of the matrix residues to diagonal form. E.g., let us execute the following command
\lstset{basicstyle={\small\sffamily}}
\begin{mmaCell}[moredefined={t, JDSpace, ds1}]{Input}
    t=Transpose[JDSpace[\mmaFrac{SeriesCoefficient[ds1[x],\{x,0,-1\}]}{\mmaUnd{\(\pmb{\epsilon}\)}}]]
\end{mmaCell}
The inner \mmaInlineCell{Input}{SeriesCoefficient[...]} calculates the matrix residue at $x=0$. We divide it by $\e$ to avoid $\e$-dependence. The function \mmaInlineCell[moredefined=JDSpace]{Input}{JDSpace[m]} (``\mmaInlineCell[moredefined=JD]{Input}{JD}'' stands for ``Jordan Decomposition'') finds the list of generalized eigenvectors of the matrix \mmaInlineCell[moredefined=JDSpace]{Input}{m}. The transposition gives the transformation to the corresponding basis. Finally,
\begin{mmaCell}[moredefined={Transform,ds1,t}]{Input}
    Transform[ds1,t]
\end{mmaCell}
gives a somewhat simpler form
\lstset{basicstyle={\tiny\sffamily}}
\begin{mmaCell}{Output}
    \{\{-\mmaFrac{2 (\(\epsilon\)+2 x \(\epsilon\)+3 \mmaSup{x}{2} \(\epsilon\))}{(x-1) x (1+x)},0,\mmaFrac{14 (-2 \(\epsilon\)+x \(\epsilon\))}{9 (x-1) (1+x)},-\mmaFrac{-14 \(\epsilon\)}{(x-1) (1+x)},0\},\{\mmaFrac{-18 (\(\epsilon\)+3 x \(\epsilon\))}{7 (x-1) (1+x)},\mmaFrac{4 x \(\epsilon\)}{(x-1) (1+x)},\mmaFrac{-\(\epsilon\)-2 x \(\epsilon\)+2 \mmaSup{x}{2} \(\epsilon\)}{(x-1) x (1+x)},\mmaFrac{-9 \(\epsilon\)}{(x-1) (1+x)},0\},
    \{0,-\mmaFrac{4 (\(\epsilon\)+3 x \(\epsilon\))}{(x-1) (1+x)},-\mmaFrac{6 \(\epsilon\)}{(x-1) (1+x)},-\mmaFrac{18 \(\epsilon\)}{(x-1) (1+x)},0\},\{-\mmaFrac{32 \(\epsilon\)}{7 (x-1) (1+x)},\mmaFrac{4 (5 \(\epsilon\)+7 x \(\epsilon\))}{9 (x-1) (1+x)},\mmaFrac{26 \(\epsilon\)}{9 (x-1) (1+x)},\mmaFrac{6 \(\epsilon\)}{(x-1) (1+x)},0\},
    \{-\mmaFrac{48 \(\epsilon\)}{7 (x-1) (1+x)},-\mmaFrac{4 \(\epsilon\)}{3 (1+x)},\mmaFrac{10 \(\epsilon\)}{3 (x-1) (1+x)},\mmaFrac{6 \(\epsilon\)}{(x-1) (1+x)},-\mmaFrac{4 \(\epsilon\)}{(x-1) (1+x)}\}\}
\end{mmaCell}
If it is not that obvious that $\e$ factors out, we can check it with
\lstset{basicstyle={\small\sffamily}}
\begin{mmaCell}[moredefined={EFormQ,ds1}]{Input}
    EFormQ[ds1,\mmaUnd{\(\pmb{\epsilon}\)}]
\end{mmaCell}
\begin{mmaCell}{Output}
    True
\end{mmaCell}
Let us finally gather all our work in association list with
\begin{mmaCell}[moredefined={OverallTransformation,ds1}]{Input}
    transformation=OverallTransformation[ds1];
\end{mmaCell}
Now we can retrieve all necessary information from various fields of  \mmaInlineCell[moredefined=transformation]{Input}{transformation}. The code below should demonstrate the most important fields:
\begin{mmaCell}[moredefined={transformation,Transform,m,ds1}]{Input}
    Mi=transformation[In][x];(*initial matrix*)
    Mf=transformation[Out][x];(*transformed matrix*)
    T=transformation[Transform];(*overall transformation matrix*)
    (*Check that everything is Ok*)
    Mi == m \&\& Mf == ds1[x] \&\& Factor[Transform[m, T, x]] == Factor[Mf]
\end{mmaCell}
\begin{mmaCell}{Output}
    True
\end{mmaCell}
We can now save our work with \mmaInlineCell[moredefined=transformation]{Input}{transformation>\!\!>"transformation.m"}.

%% file: example2.tex

\subsection{Example 2: energy loss in electron-nucleus bremsstrahlung.}

Let us now preset a full-fledged example of using \Libra in physical applications. It will allow us to demonstrate the use of a few functions which did not appear in the previous example.

We will calculate the electron energy loss in the process of Bremsstrahlung on the nucleus. I.e., we will rederive Racah result \cite{Racah1934} for the photon-energy weighted cross section
\begin{equation}
    \phi_{\text{rad}} = \int \frac{\omega}{\varepsilon} d\sigma_{eZ\to eZ\gamma}\,,
\end{equation}
where $\omega$ is the photon energy and $\varepsilon$ is the energy of the initial electron. From now on we will put electron mass to unity.

\begin{figure}
    \centering
    \includegraphics[width=1\textwidth]{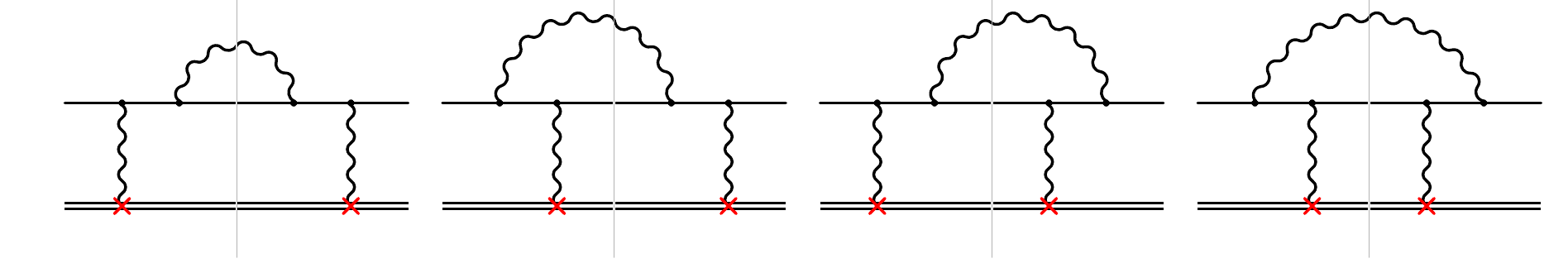}
    \caption[BS energy loss]{Diagrams contributing to the cross section of Bremsstrahlung.}
    \label{fig:BS}
\end{figure}

Using Cutkosky rules, we express the energy-weighted cross section via cut diagrams shown in Fig. \ref{fig:BS} with the integrand multiplied by $\omega$. 
We define the following family of integrals:
\begin{equation}
    j_{n_1,\ldots,n_7}=e^{2\e \gamma_E}\int \frac{d^dp_2 d^dk_1}{\pi^d}\frac{\prod_{k=1}^{3}\delta^{(n_k-1)}(-D_k)}{D_4^{n_4}D_5^{n_5}D_6^{n_6}D_7^{n_7}}
\end{equation}
where
\begin{gather}
    D_1=n\cdot \left(k_1-p_1+p_2\right),\ 
    D_2=p_2^2-1,\ 
    D_3 = k_1^2,\
    D_4=\left(k_1-p_1+p_2\right)^2,\nonumber\\
    D_5=\left(k_1+p_2\right)^2-1,\
    D_6=\left(p_1-k_1\right)^2-1,\
    D_7=k_1\cdot n\,,
\end{gather}
$p_1$, $p_2$, and $k_1$ are the momenta of initial electron, final electron, and final photon, respectively, and $n=(1,\boldsymbol{0})$ is the time direction. The last index, $n_7$, can not be positive.
We find the following 5 master integrals
\begin{equation}\label{eq:LaportaMasters}
    \bs j=(j_{1110000},j_{2110000},j_{1110010},j_{1110020},j_{1111000})^\intercal,
\end{equation}
which obey the differential system 
\begin{equation}
    \partial_\varepsilon \bs j = M \bs j
\end{equation}
with
\begin{equation}
M=\left(
\begin{array}{ccccc}
    0 & 1 & 0 & 0 & 0 \\
    \frac{(1 - 2 \epsilon) (4 \epsilon-3)}{\varepsilon^2-1} & 
    \frac{3 (1-2 \epsilon) \varepsilon}{\varepsilon^2-1} & 0 & 0 & 0 \\
    \frac{(1-2 \epsilon) (4 \epsilon -3)}{4 \epsilon 
        \varepsilon  \left(\varepsilon ^2-1\right)} &
    \frac{3-8 \epsilon}{4 \epsilon\left(\varepsilon ^2-1\right)} & 
    \frac{(2\epsilon -1) \varepsilon }{\varepsilon ^2-1} &
    \frac{2}{\varepsilon } & 0 \\
    \frac{(1-2 \epsilon) (3-4 \epsilon)}{8 \epsilon\varepsilon  \left(\varepsilon ^2-1\right)^2} &
    \frac{(2 \epsilon -1) \left(4 \epsilon 
        \varepsilon ^2-4 \epsilon +3\right)}{8 \epsilon 
        \left(\varepsilon ^2-1\right)^2} & 
    \frac{2\epsilon  (1-2 \epsilon) \varepsilon}{\varepsilon ^2-1} & 
    -\frac{4 \epsilon \varepsilon ^2-\varepsilon ^2+1}{\varepsilon \left(\varepsilon ^2-1\right)} & 0 \\
    0 & -\frac{1}{\varepsilon ^2-1} & 0 & 0 & \frac{(2\epsilon -1) \varepsilon }{\varepsilon ^2-1} \\
\end{array}
\right)
\end{equation}
\mmaSet{index=1}
From now on we can use \Libra.
{
\begin{mmaCell}[moredefined=Libra]{Input}
    Block[\{Print\},<\!\!<Libra\textasciigrave];
    M =
\end{mmaCell}
\vspace{-9mm}
\small
\begin{verbatim}
      {{0,1,0,0,0},{(1-2*e)*(4*e-3)/pp,3*(1-2*e)*w/pp,0,0,0},
{(1-2*e)*(4*e-3)/4/e/w/pp,(3-8*e)/4/e/pp,(2*e-1)*w/pp,2/w,0},
{(1-2*e)*(3-4*e)/8/e/w/pp^2,(2*e-1)*(3/pp+4*e)/8/e/pp,2*e*(1-2*e)*w/pp,
(1-4*e*w^2/pp)/w,0}, {0,-1/pp,0,0,(2*e-1)*w/pp}}//.{pp->w^2-1,
e->\[Epsilon],w->\[CurlyEpsilon]};
\end{verbatim}
}
\begin{mmaCell}[moredefined={NewDSystem,M}]{Input}
    NewDSystem[bsde,\mmaUnd{\(\pmb{\varepsilon}\)}\(\pmb{\to}\)M];
\end{mmaCell}
The matrix $M$ is block-triangular, and we can use the following command to discover the indices of the diagonal blocks:
\begin{mmaCell}[moredefined={EntangledBlocksIndices,M,bsde}]{Input}
    EntangledBlocksIndices[bsde]
\end{mmaCell}
\begin{mmaCell}{Output}
    \{\{1, 2\}, \{3, 4\}, \{5\}\}
\end{mmaCell}

First, we have to reduce the diagonal blocks.
We start from the block \texttt{\{1,2\}}. The basic strategy is to copy the block under consideration into temporary variable. Then we can transform this block with a sequence of transformations and apply to the whole matrix the overall transformation.
\begin{mmaCell}[moredefined={NewDSystem,M,bsde}]{Input}
    ii=\{1, 2\};
    NewDSystem[b, \mmaUnd{\(\pmb{\varepsilon}\)}\(\pmb{\to}\)bsde[[ii,ii]]];
\end{mmaCell}
Now we run the visual tool
\begin{mmaCell}[moredefined={VisTransformation,b,M,bsde}]{Input}
    t = VisTransformation[b, \mmaUnd{\(\pmb{\varepsilon}\)}, \mmaUnd{\(\pmb{\epsilon}\)}];
\end{mmaCell}
and immediately see the problem:
\begin{center}
    \includegraphics[width=0.5\linewidth]{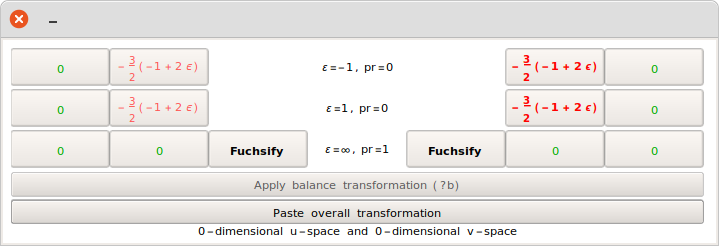}
\end{center}
The eigenvalues of the matrix residues at $\varepsilon=\pm 1$ are half-integer at $\e=0$. Therefore, we have to make variable change. Following the receipt of Ref. \cite{Lee2017c}, we find the appropriate variable change:
\begin{equation}
    \varepsilon = \frac{1+z^2}{1-z^2}\,.
\end{equation}
This variable changes from $0$ to $1$ when $\varepsilon$ increases from $1$ to $\infty$.

There are two different ways to introduce a new variable in \Libra. The first one is based on a global variable change. This method has obvious limitations in the case when it is not possible to make the global rationalizing variable change. Another method is based on the command \mmaInlineCell[moredefined=AddNotations]{Input}{AddNotations}. This method is more involved, and we refer to tutorials which come with \Libra distribution. For our present example it is sufficient to make the global variable change. So we execute the command
\begin{mmaCell}[moredefined={ChangeVar,bsde}]{Input}
    ChangeVar[bsde,\mmaUnd{\(\pmb{\varepsilon}\)}\(\pmb{\to}\)(1+z^2)/(1-z^2),z];
\end{mmaCell}

\subsubsection*{Reducing block $\{1, 2\}$.}
Next, we return to the reduction of the of the \texttt{\{1,2\}} block. We reinitialize the temporary matrix
\begin{mmaCell}[moredefined={NewDSystem,VisTransformation,bsde}]{Input}
    NewDSystem[b, \mmaUnd{\(\pmb{\varepsilon}\)}\(\pmb{\to}\)bsde[[ii,ii]]];
\end{mmaCell}
and run the visual tool. Note that the second argument of \mmaInlineCell[moredefined=VisTransformation]{Input}{VisTransformation} should now be $z$ rather than $\varepsilon$. To save space, instead of drawing intermediate pictures of the visual tool, we will indicate the option \mmaInlineCell[moredefined=Highlighted]{Input}{Highlighted->...} to guide the reader by highlighting the buttons to be pressed. 
\begin{mmaCell}[moredefined={b,t,Highlighted,VisTransformation,Transform}]{Input}
    t = VisTransformation[b, z, \mmaUnd{\(\pmb{\epsilon}\)}, Highlighted \(\pmb{\to}\)
            \{\{3\}\(\pmb{\leftrightarrow}\)\{5\},\{7\}\(\pmb{\leftrightarrow}\)\{9\},\{1\}\(\pmb{\leftrightarrow}\)\{4\},\{5\}\(\pmb{\leftrightarrow}\)\{8\},\{1\}\(\pmb{\leftrightarrow}\)\{4\},\{5\}\(\pmb{\leftrightarrow}\)\{8\}\}];
    Transform[b,t];
\end{mmaCell}
Finally, we factor out $\e$ dependence and diagonalize the matrix residue at $z=0$ similarly to the first example:
\begin{mmaCell}[moredefined={b,t,FactorOut,JDSpace,Transform}]{Input}
    t = FactorOut[b, z, \mmaUnd{\(\pmb{\epsilon}\)}], 1/2] /. \_C \(\pmb{\to}\) 1;
    Transform[b,t];
    t = Transpose@JDSpace[SeriesCoefficient[b[z], {z, 0, -1}]/ \mmaUnd{\(\pmb{\epsilon}\)}];
    Transform[b,t];
    b[z]
\end{mmaCell}
\begin{mmaCell}{Output}
    \{\{0,-\mmaFrac{24 \(\epsilon\)}{(-1+z) (1+z)}\},\{\mmaFrac{4 \(\epsilon\)}{3 (-1+z) (1+z)},\mmaFrac{6 (\(\epsilon\)+\mmaSup{z}{2} \(\epsilon\))}{(-1+z) z (1+z)}\}\}
\end{mmaCell}
Now we consolidate the sequence of transformations made into one transformation and apply it to the big system \mmaInlineCell[moredefined={bsde}]{Input}{bsde}:
\begin{mmaCell}[moredefined={b,t,ii,bsde,HistoryConsolidate,Transform}]{Input}
    t = HistoryConsolidate[b];(*shortcut for OverallTransformation[b][Transform]*)  
    Transform[bsde,t,ii];
\end{mmaCell}
Note that third argument of \mmaInlineCell[moredefined={Transform}]{Input}{Transform} now indicates the indices of the block \mmaInlineCell[moredefined={ii}]{Input}{ii=\{1, 2\}}. We can check that the block is in $\e$-form
\begin{mmaCell}[moredefined={b,t,ii,bsde,EFormQ}]{Input}
    EFormQ[bsde[[ii, ii]],\mmaUnd{\(\pmb{\epsilon}\)}]
\end{mmaCell}
\begin{mmaCell}{Output}
    True
\end{mmaCell}
\subsubsection*{Reducing block $\{3, 4\}$.}
Another $2\times2$ block $\{3,4\}$ is reduced in a similar way:
\begin{mmaCell}[moredefined={b,t,ii,z,bsde,NewDSystem,VisTransformation,Animate,Transform,FactorOut,JDSpace,HistoryConsolidate,EFormQ}]{Input}
    ii = \{3, 4\};
    NewDSystem[b, z\(\pmb{\to}\)bsde[[ii,ii]]];
    t = VisTransformation[b, z, \mmaUnd{\(\pmb{\epsilon}\)}, Animate \(\pmb{\to}\)(*Animate will press buttons 4 U*)
    \{\{2\}\(\pmb{\leftrightarrow}\)\{5\},\{8\}\(\pmb{\leftrightarrow}\)\{5\},\{3\}\(\pmb{\leftrightarrow}\)\{1\},\{7\}\(\pmb{\leftrightarrow}\)\{5\}\}];
    Transform[b, t];
    t = FactorOut[b, z, \mmaUnd{\(\pmb{\epsilon}\)}, -1/2] /.\_C \(\pmb{\to}\) 1;
    Transform[b, t];
    t = Transpose@JDSpace[SeriesCoefficient[b[z], \{z, 0, -1\}]/ \mmaUnd{\(\pmb{\epsilon}\)}];
    Transform[b, t];
    t = HistoryConsolidate[b];  
    Transform[bsde,t,ii];
    EFormQ[bsde,\mmaUnd{\(\pmb{\epsilon}\)}]
\end{mmaCell}
\begin{mmaCell}{Output}
    True
\end{mmaCell}

\subsubsection*{Reducing block $\{5\}$.}
Finally, the last $1\times1$ block can be reduced either using the same technique, or by explicit integration:
\begin{mmaCell}[moredefined={b,t,ii,z,bsde,EFormQ,Transform}]{Input}
    t = \{\{Exp[Integrate[Factor[bsde[z][[5, 5]]] /. \mmaUnd{\(\pmb{\epsilon}\)} \(\pmb{\to}\) 0, z]]\}\};
    Transform[bsde, t, {5}];
    EFormQ[bsde[[{5}, {5}]], \mmaUnd{\(\pmb{\epsilon}\)}]
\end{mmaCell}
\begin{mmaCell}{Output}
    True
\end{mmaCell}

\subsubsection*{Reducing off-diagonal blocks.}
At this stage we have reduced all diagonal blocks. Now we have to take care of the off-diagonal blocks. First, we get rid of the multiple poles with \mmaInlineCell[moredefined={Fuchsify}]{Input}{Fuchsify} command:
\begin{mmaCell}[moredefined={b,t,ii,z,bsde,EFormQ,Transform,Fuchsify,FuchsianQ}]{Input}
    t = Fuchsify[bsde,z];
    Transform[bsde, t];
    FuchsianQ[bsde,z]
\end{mmaCell}
\begin{mmaCell}{Output}
    True
\end{mmaCell}

\subsubsection*{Factoring out $\e$ from the whole matrix.}
Finally, we factor out $\e$ from the whole matrix:
\begin{mmaCell}[moredefined={b,t,ii,z,bsde,EFormQ,Transform,FactorOut}]{Input}
    t = FactorOut[bsde, z, \mmaUnd{\(\pmb{\epsilon}\)}, 1] /. \_C  \(\pmb{\to}\)  1
    Transform[bsde, t];
    EFormQ[bsde, \mmaUnd{\(\pmb{\epsilon}\)}]
\end{mmaCell}
\begin{mmaCell}{Output}
    True
\end{mmaCell}

\subsubsection*{Gathering data.}
We have obtained the $\e$-form and it is a perfect time to save our work with
\begin{mmaCell}[moredefined={b,t,ii,z,bsde,EFormQ,Transform,OverallTransformation}]{Input}
    transformation = OverallTransformation[bsde];
    Put[transformation, "transformation"];
    Quit
\end{mmaCell}
and to take a cup of coffee.

\subsubsection{Fixing boundary conditions.}
\mmaSet{index=1}
Ok, after a short break, we return to our calculations. We load \Libra and our previous work with
\begin{mmaCell}[moredefined=Libra]{Input}
    Block[\{Print\},<\!\!<Libra\textasciigrave];
    transformation = Get["transformation"];
    Mf = transformation[Out][z];(*final matrix in epsilon-form*)
    T = transformation[Transform];(*transformation matrix*)
\end{mmaCell}

A nice feature of \Libra is that it can help one in fixing the boundary conditions. Namely, it can determine the minimal set of coefficients in the asymptotics of the Laporta master integrals which are to be calculated. For this purpose one should use the \mmaInlineCell[moredefined={GetLcs}]{Input}{GetLcs} command. This command returns two objects: a matrix $L$ and the list of coefficients $\bs c$ (thus the name of the command). Let us run this command for our example. 

We want to fix boundary conditions from the threshold asymptotics, thus, at $z\to0$. Note that $z=0$ is a singular point of our system. Therefore, for the demonstration purpose, let us first pretend that we want to put boundary conditions at some regular point $z=z_0$. For this case we simply have to fix the values of the Laporta master integrals at $z=z_0$. The values of the canonical masters at $z=z_0$ are determined by the obvious formula
\begin{equation}\label{eq:Lcsz0}
    \bs J(z_0)= T^{-1}(z_0)\bs j(z_0)
\end{equation}
Let us now see how the same conclusions follow from the execution of \mmaInlineCell[moredefined={GetLcs}]{Input}{GetLcs} command. Let us put $z_0=1/2$ for example. Then we execute 
\begin{mmaCell}[moredefined={Mf,T,bsde,GetLcs,Transform,OverallTransformation}]{Input}
    With[\{z0 = 1/2\}, 
        \{L,cs\} = GetLcs[Mf,T,\{z, z0\}]
    ];
\end{mmaCell}
The meaning of arguments of \mmaInlineCell[moredefined={GetLcs}]{Input}{GetLcs} should be self-explaining. We only remark that the argument \mmaInlineCell{Input}{\{z, z0\}} can be replaced with \mmaInlineCell{Input}{\{z, z - z0\}} without changing the result of the program\footnote{One should not worry about the apparent ambiguity. To disambiguate the two syntaxes, \Libra simply checks if the second element of the pair depends on the first element.}. We will explain below why the second syntax is more expressive.

Let us now examine the list of coefficients:
\begin{mmaCell}[moredefined={Mf,T,bsde,GetLcs,Transform,OverallTransformation}]{Input}
    cs
\end{mmaCell}
\begin{mmaCell}{Output}
    \{\{1, 0, 0\}, \{2, 0, 0\}, \{3, 0, 0\}, \{4, 0, 0\}, \{5, 0, 0\}\}
\end{mmaCell}
This is the list of triples, the triple $\{i,\alpha,k\}$ denotes the coefficient in front of $(z-z_0)^\alpha \ln^{k}(z-z_0)$ in $z\to z_0$ asymptotics of $i$-th Laporta integral. Therefore, the list above suggests us to calculate the value of each Laporta integral at $z=z_0$, exactly as we have anticipated. Next, the matrix $L$ is the ``adapter'' between the coefficients of Laporta master integrals and the column of boundary constants for canonical master integrals. According to Eq. \eqref{eq:Lcsz0}, we expect that $L=T^{-1}(z_0)$. Let us check that this is indeed the case:
\begin{mmaCell}[moredefined={Mf,T,L,bsde,GetLcs,Transform,OverallTransformation}]{Input}
    Factor[L] == Factor[Inverse[T /. {z -> 1/2}]]
\end{mmaCell}
\begin{mmaCell}{Output}
    True
\end{mmaCell}

So, if we always wanted to put boundary condition at regular points, there would be no need for the dedicated procedure \mmaInlineCell[moredefined={GetLcs}]{Input}{GetLcs}. However, the problem is that we usually want to fix boundary conditions from the asymptotics at some singular point of the differential system. In particular, in our example we want to fix boundary conditions from threshold asymptotics $z\to 0$. This is where  \mmaInlineCell[moredefined={GetLcs}]{Input}{GetLcs} becomes really helpful. So, let us execute
\begin{mmaCell}[moredefined={L,cs,Mf,T,bsde,GetLcs,Transform,OverallTransformation}]{Input}
    depth = 3;preferred=1|3|5;
    \{L,cs\} = GetLcs[Mf, T, \{z, z, depth\}, preferred];
\end{mmaCell}
Note, that \mmaInlineCell[moredefined={L,cs,Mf,T,GetLcs}]{Input}{\{L,cs\} = GetLcs[Mf,T,\{z,z\}];} would also work, but would provide a slightly different set of constants. Of course, there is no wonder that we can fix boundary conditions from different sets of asymptotic coefficients, but some constants might be more approachable for calculation. \Libra gives a user several ways to provide a hint of what constants (s)he prefers to calculate. Some of them are demonstrated in the command above. First, the optional parameter \mmaInlineCell[moredefined=preferred]{Input}{preferred} tells that we prefer to calculate the asymptotics of the Laporta integrals \#\#1,3,5 from the list in Eq. \eqref{eq:LaportaMasters}. Next, the parameter  \mmaInlineCell[moredefined=depth]{Input}{depth} defines the extra depth of the search that \Libra does.

Let us now examine the list of constants that we have to calculate:
\begin{mmaCell}[moredefined={Mf,T,bsde,GetLcs,Transform,OverallTransformation}]{Input}
    cs
\end{mmaCell}
\begin{mmaCell}{Output}
    \{\{1, 0, 0\}, \{1, 5 - 6 \(\epsilon\), 0\}, \{3, 2 - 4 \(\epsilon\),0\}, \{3, -1 + 2 \(\epsilon\), 0\}, \{5, -1 + 2 \(\epsilon\), 0\}\}
\end{mmaCell}
Note that now we have fractional powers\footnote{This is the reason why the syntax \texttt{\{z, z - z0\}} is preferred over \texttt{\{z, z0\}} in the third argument of  \texttt{GetLcs} procedure. It gives us a way to explicitly indicate which variable is to be used in the expansion. For example, \texttt{\{z, 1-z\}} means that we want to expand in powers of $(1-z)$ rather than those of $(z-1)$. The difference is, of course, only the complex phase, but using the advanced syntax, one can avoid unnecessary errors in the definition of this phase.}.

In order to find the required constants, one can use the expansion by regions method. In fact, there are simple considerations (falling beyond the scope of the present paper) which tell us that only constant $\{1, 5 - 6 \epsilon, 0\}$ can be nonzero and should be explicitly calculated. This constant corresponds to the coefficient in the leading threshold asymptotics of the phase-space integral. Simple calculation results in
\begin{equation}\label{eq:asy}
    j_{1110000}\stackrel{z\to 0}{\sim}  e^{2\e \gamma_E}\frac{2^{4-6 \epsilon } \Gamma (1-\epsilon )}{\pi ^{3/2} \Gamma \left(\frac{7}{2}-3 \epsilon \right)} z^{5-6\e}
\end{equation}

Therefore, we define a list of calculated constants
\begin{mmaCell}[moredefined={ cs}]{Input}
    csvals=Replace[cs,\{\{1,5-6 \mmaUnd{\(\pmb{\epsilon}\)},0\}\(\pmb{\to}\)\mmaFrac{\mmaSup{2}{4-6 \mmaUnd{\(\pmb{\epsilon}\)}} Gamma[1-\mmaUnd{\(\pmb{\epsilon}\)}]}{\mmaSup{\mmaDef{\(\pmb{\pi}\)}}{3/2} Gamma[\mmaFrac{7}{2}-3 \mmaUnd{\(\pmb{\epsilon}\)}]}Exp[2\mmaUnd{\(\pmb{\epsilon}\)} EulerGamma], 
                _\(\pmb{\to}\)0\},\{1\}]
\end{mmaCell}
\begin{mmaCell}{Output}
    \{0,Exp[2\(\epsilon\) EulerGamma]\mmaFrac{\mmaSup{2}{4-6 \(\epsilon\)} Gamma[1-\(\epsilon\)]}{\mmaSup{\(\pi\)}{3/2} Gamma[\mmaFrac{7}{2}-3 \(\epsilon\)]},0,0,0\}
\end{mmaCell}

\subsubsection{Constructing solution for canonical master integrals.}

At this point we have determined all necessary ingredients for constructing the solution for the canonical master integrals in terms of iterated integrals. First, we construct the general solution
\begin{mmaCell}[moredefined={Mf,T,bsde,GetLcs,Transform,PexpExpansion}]{Input}
    o = 4;(*maximal order of epsilon expansion*)
    U = PexpExpansion[\{Mf,o\}, z, Split->False] + O[\mmaUnd{\(\pmb{\epsilon}\)}]^(o+1);
    Dimensions[U]
\end{mmaCell}
\begin{mmaCell}{Output}
    \{5, 5\}
\end{mmaCell}
The code above should be self-explaining, except the \mmaInlineCell[moredefined=depth]{Input}{Split}
option which determines whether to split successive terms of expansion or to sum them up.
Note that we don't need to explicitly indicate the lower limit of integration (the point where we put our boundary conditions) in the path-ordered exponent, we just indicate the variable ($z$). The result is given in terms of abstract iterated integrals \texttt{II} whose lower limit is implied to be the point where we put our boundary conditions. For our case, it is simply the point $z=0$, so we can safely replace \texttt{II} with \texttt{G} to obtain the Goncharov's polylogarithms as they are defined, e.g., in \texttt{GinaC}, \cite{bauer2012ginac}. So we do
\begin{mmaCell}[moredefined={U,II}]{Input}
    U = U/.\{II[\{\},_] \(\pmb{\to}\) 1, II \(\pmb{\to}\) G\};
\end{mmaCell}

Finally, we obtain the expansion for the canonical master integrals as a matrix product
\begin{mmaCell}[moredefined={U,L,csvals}]{Input}
    Csvals =L.csvals;
    Jsvals = U.Csvals;
\end{mmaCell}
Note that we prefer here first to multiply the exact matrices  \mmaInlineCell[moredefined=L]{Input}{L} and \mmaInlineCell[moredefined=csvals]{Input}{csvals} and then to multiply the result by the expansion \mmaInlineCell[moredefined=U]{Input}{U}. This is the rule of thumb which prevents one from losing orders in $\e$-expansion.

Our results for \mmaInlineCell[moredefined=Jsvals]{Input}{Jsvals} are already good for using in physical application, but there is a little defect in them that we are going to fix now.
Namely, if we examine carefully the obtained expansion of the canonical integrals, we will see that it lacks the property of uniform transcendentality. The explanation is simple: the transformation matrix $T$ to $\e$-form is defined with some freedom. At least, we can multiply it by some factor $f(\e)$ independent of $z$.  For our present case, when there is only one non-zero boundary constant defined in Eq. \eqref{eq:asy}, we can derive $f$ in a straight-forward way.\footnote{In more complicated cases there is a simple empirical approach allowing to find $f$ provided that we know sufficiently many terms of $\e$-expansion of boundary constants. This approach goes beyond the scope of the present paper.} First, we note that $U$ is uniform transcendental by construction. Therefore, we examine carefully any nonzero entry of  $\bs C =$\mmaInlineCell[moredefined=Csvals]{Input}{Csvals}. E.g., we have 
\begin{equation}
    C_2 =-\frac{9\cdot 2^{-6 \epsilon} e^{2  \epsilon  \gamma_E} \Gamma (-\epsilon )}{4\pi ^{3/2} (4 \epsilon -3) (4 \epsilon -1) \Gamma \left(\frac{1}{2}-3 \epsilon \right)}
\end{equation}
Now we transform this expression bearing in mind that, for $k\in \mathbb{Z}$ and $r\in \mathbb{Q}$, the expressions 
\begin{equation*}
    r,\ r^{k \e}, \text{ and }\tilde{\Gamma}(1+k\e)\stackrel{\text{def}}{=}e^{k \epsilon  \gamma_E} \Gamma(1+k\e)
\end{equation*}
have uniform $\e$-expansions. We have
\begin{equation}
    C_2 =\frac{9}{4 \pi ^2 \epsilon  (4 \epsilon -3) (4 \epsilon -1)} \times \frac{2^{-12 \epsilon }\tilde{\Gamma} (1-3 \epsilon ) \tilde{\Gamma} (1-\epsilon )}{\tilde{\Gamma} (1-6 \epsilon )}\,.
\end{equation}
The second factor has uniform $\e$-expansion, so we can  take $f=\frac{9}{4 \pi ^2 \epsilon  (4 \epsilon -3) (4 \epsilon -1)}$. Now we have to redefine $T\to T\,f$, $L\to L/f$.

Let us apply this fix to our data (with due precautions!):
\begin{mmaCell}[moredefined={T,U,L,Csvals,csvals,Jsvals,transformation,Transform,II}]{Input}
    f=9/4/Pi^2/\(\pmb{\epsilon}\)/(3 - 4*\(\pmb{\epsilon}\))/(1 - 4*\(\pmb{\epsilon}\));
    T1 = T*f; L1 = L/f;
    transformation1 = transformation;
    transformation1[Transform] = T1;
\end{mmaCell}
To avoid mistakes at this stage, we have created new temporary variables \mmaInlineCell[moredefined=T1]{Input}{T1}, \mmaInlineCell[moredefined=L1]{Input}{L1}, and \mmaInlineCell[moredefined=transformation1]{Input}{transformation1}. Now, before redefining the previous variables, we should better check the consistency. First, we check that new transformation matrix leads to the same final matrix:
\begin{mmaCell}[moredefined={T,U,L,Csvals,csvals,Jsvals,transformation1,Mf,Transform,ChangeVar}]{Input}
    Mi=ChangeVar[transformation1[In][\mmaUnd{\(\pmb{\varepsilon}\)}], \mmaUnd{\(\pmb{\varepsilon}\)} \(\pmb{\to}\) (1+z^2)/(1-z^2),z];
    Factor[Transform[Mi,transformation1[Transform],z]-Mf]
\end{mmaCell}
\begin{mmaCell}{Output}
    \{\{0,0,0,0,0\},\{0,0,0,0,0\},\{0,0,0,0,0\},\{0,0,0,0,0\},\{0,0,0,0,0\}\}
\end{mmaCell}
Next, we should check if the matrix $L$ is defined correctly. For this purpose we can use the function \mmaInlineCell[moredefined=GetL]{Input}{GetL}, which admits as the fourth parameter the list of coefficients and constructs the corresponding adapter matrix:
\begin{mmaCell}[moredefined={T1,U,L1,cs,Mf,Transform,GetL}]{Input}
    Factor[GetL[Mf, T1, \{z, z\}, cs] - L1]
\end{mmaCell}
\begin{mmaCell}{Output}
    \{\{0,0,0,0,0\},\{0,0,0,0,0\},\{0,0,0,0,0\},\{0,0,0,0,0\},\{0,0,0,0,0\}\}
\end{mmaCell}
Since the data seems to be consistent, we reassign the old variables and remember to update the file:
\begin{mmaCell}[moredefined={T,T1,U,L,L1,Csvals,csvals,Jsvals,transformation,transformation1,Transform,II}]{Input}
    T=T1;L=L1;transformation=transformation1;
    Put[transformation, "transformation"];    
    Csvals = L.csvals;
    Jsvals =ExpandAll@FunctionExpand[U.Csvals];
\end{mmaCell}
Let us examine the obtained results by an eye by printing only two leading terms of the $\e$ expansion\footnote{\texttt{LeadingSeries} probably deserves to be an inherent \textit{Mathematica} function, but for now it is a function defined in \Libra.}. We note the proliferation of $\ln 2$, so to save space we present here the expansion of  \mmaInlineCell[moredefined=Jsvals]{Input}{(\mmaSup{2}{12 \mmaUnd{\(\pmb{\epsilon}\)}}Jsvals)} which appears to be free of $\ln 2$:
\begin{mmaCell}[moredefined={LeadingSeries, Jsvals},morepattern={\#}]{Input}
    Simplify[LeadingSeries[#,\{\mmaUnd{\(\pmb{\epsilon}\)},0,1\}]]&/@(\mmaSup{2}{12 \mmaUnd{\(\pmb{\epsilon}\)}}Jsvals)
\end{mmaCell}

\lstset{basicstyle={\tiny\sffamily}}
\begin{mmaCell}{Output}
    \{(12 G[\{-1\},z]-12 G[\{1\},z]) \(\epsilon\)+72 (G[\{-1,-1\},z]-G[\{-1,0\},z]+G[\{-1,1\},z]-G[\{1,-1\},z]+G[\{1,0\},z]-G[\{1,1\},z]) \mmaSup{\(\epsilon\)}{2}+\mmaSup{O[\(\epsilon\)]}{3},
    1+6 (G[\{-1\},z]-G[\{0\},z]+G[\{1\},z]) \(\epsilon\)+\mmaSup{O[\(\epsilon\)]}{2},\mmaFrac{4}{3} (G[\{-1,-1\},z]-G[\{-1,1\},z]-G[\{1,-1\},z]+G[\{1,1\},z]) \mmaSup{\(\epsilon\)}{2}+
    \mmaFrac{8}{3} (3 G[\{-1,-1,-1\},z]-3 G[\{-1,-1,0\},z]+3 G[\{-1,-1,1\},z]-G[\{-1,0,-1\},z]+G[\{-1,0,1\},z]-G[\{-1,1,-1\},z]+
    3 G[\{-1,1,0\},z]-5 G[\{-1,1,1\},z]+G[\{0,-1,-1\},z]-G[\{0,-1,1\},z]-G[\{0,1,-1\},z]+G[\{0,1,1\},z]-5 G[\{1,-1,-1\},z]+
    3 G[\{1,-1,0\},z]-G[\{1,-1,1\},z]+G[\{1,0,-1\},z]-G[\{1,0,1\},z]+3 G[\{1,1,-1\},z]-3 G[\{1,1,0\},z]+3 G[\{1,1,1\},z]) \mmaSup{\(\epsilon\)}{3}+\mmaSup{O[\(\epsilon\)]}{4},
    \mmaFrac{8}{27} (G[\{-1\},z]-G[\{1\},z]) \(\epsilon\)+\mmaFrac{4}{9} (5 G[\{-1,-1\},z]-4 G[\{-1,0\},z]+3 G[\{-1,1\},z]-G[\{0,-1\},z]+G[\{0,1\},z]-3 G[\{1,-1\},z]+
    4 G[\{1,0\},z]-5 G[\{1,1\},z]) \mmaSup{\(\epsilon\)}{2}+\mmaSup{O[\(\epsilon\)]}{3},-2 (G[\{-1,-1\},z]-G[\{-1,1\},z]-G[\{1,-1\},z]+G[\{1,1\},z]) \mmaSup{\(\epsilon\)}{2}-4 (2 G[\{-1,-1,-1\},z]-
    3 G[\{-1,-1,0\},z]+4 G[\{-1,-1,1\},z]-2 G[\{-1,1,-1\},z]+3 G[\{-1,1,0\},z]-4 G[\{-1,1,1\},z]+G[\{0,-1,-1\},z]-G[\{0,-1,1\},z]-
    G[\{0,1,-1\},z]+G[\{0,1,1\},z]-4 G[\{1,-1,-1\},z]+3 G[\{1,-1,0\},z]-2 G[\{1,-1,1\},z]+4 G[\{1,1,-1\},z]-3 G[\{1,1,0\},z]+
    2 G[\{1,1,1\},z]) \mmaSup{\(\epsilon\)}{3}+\mmaSup{O[\(\epsilon\)]}{4}\}
\end{mmaCell}
Indeed, we see the uniform transcendental weight.

Therefore, we have calculated all integrals needed for the calculation of the energy loss. Performing some Dirac matrix algebra an expressing the functions $G$ via classical polylogarithms, we obtain
\begin{multline}
    \phi_{\text{rad}} = \alpha (Z\alpha)^2\bigg\{
    \tfrac{8 \left(z^4+z^2+1\right) }{3
        \left(z^3+z\right)}\ln{\tfrac{1+z}{1-z}}-\tfrac{4}{3}
    -\tfrac{(2z^2+3z+2)
        \left(z^2-1\right)^2}{3
        \left(z^4+z^2\right)} \ln
    ^2{\tfrac{1+z}{1-z}}\\
    +\tfrac{2 \left(z^2-1\right)^2}{z^3
        +z}\left[\text{Li}_2\left(\tfrac{1-z}{2}\right)-\text{Li}_2\left(\tfrac{1+z}{2}\right)-\text{Li}_2(-z)+\text{Li}_2(z)-\ln{\tfrac{1-z}{2}} \ln{\tfrac{1+z}{1-z}}\right]
    \bigg\}\,,
\end{multline}
which, after a little fiddling with dilogarithms, reduces to Racah result \cite{Racah1934}.

%% file: conclusion.tex

\section{Conclusion}

In the present paper we have described a new \textit{Mathematica} package \Libra dedicated to the reduction of the differential systems, in particular,those, which appear in multiloop calculations. Although, we have presented some of \Libra's features, many have been left behind the scene. Let us list some of those
\begin{enumerate}
    \item Reducing multivariate systems (see Example 5 in the \texttt{Tutorial1.nb} notebook attached to the distribution).
    \item Many linear algebra tools (\texttt{ESpace}, \texttt{EValues}, \texttt{JDSpace},\texttt{OMatrixExp}, \texttt{ODet}, \texttt{ODot}, \texttt{JDTowers},\ldots). Some of them do the same as original \textit{Mathematica} functions, but are faster in many cases.
    \item The versions of the above functions for the matrices in the quotient ring $\mathbb{Q}[x]/p(x)$, where $p(x)$ is some irreducible polynomial (\texttt{ESpaceMod}, \texttt{EValuesMod},\ldots). Also, the corresponding versions of \texttt{Series*} functions, like \texttt{SeriesCoefficientMod}. 
    \item Treatment of irreducible denominators using \texttt{*Mod} functions.
    \item An automatic verison of \texttt{VisTransformation}, the function \texttt{Rookie} (this tool is not too advanced yet).
    \item Specialized command to construct $\sU$ and $\sV$ spaces for complicated cases (\texttt{GetSubspaces}).
    \item Converting to and from the higher-order differential equation (\texttt{ToOneDE}, \texttt{ToCompanionDS}). 
    \item Construction of generalized power series (Frobenius method, functions \texttt{SeriesSolutionData}, \texttt{ConstructSeriesSolution})
    \item Introducing new variables with notations (\texttt{AddNotation}, \texttt{Notations}, \texttt{RuleToNotation}, \texttt{NotationToRule})
    \item Using \texttt{Fermat} program \cite{lewis2008computer} via \texttt{Fermatica} interface package (R.Lee) for operations with matrices populated with rational functions (\texttt{UseFermat} option in many procedures).
    \item Implementation of the decomposition algorithm used in irreducibility criterion of Ref. \cite{Lee2017c} (\texttt{BikhoffGrothendieck} function).
    \item Transformations history manipulation functions (\texttt{HistoryCheck},\linebreak \texttt{HistoryChop},  \texttt{HistoryRecall}).
\end{enumerate}

Some of these features have experimental status, but have been used in real-life problems, including the most complicated ones. The user is encouraged to check the tutorial which come with the distribution.

\paragraph{Acknowledgments} This work has been supported by Russian Science Foundation, grant 20-12-00205. Second example stems from the collaboration with V. Bytev. I am grateful to all my collaborators (in particular, to V. Smirnov, M. Steinhauser, and A. Grozin) for the joint work on the projects in which \Libra has evolved enormously. I am also grateful to A. Pomeransky for useful discussions and interest to the work.